\documentclass[twocolumn,amsmath,amssymb,mathabx,floatfix,pra,showpacs]{revtex4}

\usepackage{graphicx}           
\usepackage{ifpdf}

\def\Rb87{^{87}\rm{Rb}}                     
\def\Li6{^{6}\rm{Li}}                       

\begin{document}

\title{Systematic Shifts for Ytterbium-ion Optical Frequency Standards}

\author{N.~Batra}
\affiliation{CSIR-National Physical Laboratory, Dr. K. S. Krishnan
Marg, New Delhi - 110012, India.}
\author{Sukhjit Singh}
\affiliation{Department of Physics, Guru Nanak Dev University,
Amritsar, Punjab-143005, India.}

\author{Amisha Arora}
\affiliation{Department of Physics, Guru Nanak Dev University,
Amritsar, Punjab-143005, India.}
\author{Bindiya Arora}
\affiliation{Department of Physics, Guru Nanak Dev University,
Amritsar, Punjab-143005, India.}
\author{S.~De}
\affiliation{CSIR-National Physical Laboratory, Dr. K. S. Krishnan
Marg, New Delhi - 110012, India.}
\author{A.~Sen~Gupta}
\affiliation{CSIR-National Physical Laboratory, Dr. K. S. Krishnan
Marg, New Delhi - 110012, India.}
%
%
\begin{abstract}
The projected systematic uncertainties of single trapped
Ytterbium-ion optical frequency standards are estimated for the
quadrupole and octupole transitions which are at wavelengths 435.5
nm and 467 nm, respectively. Finite temperature of the ion and its
interaction with the external fields introduce drift in the
measured frequency compared to its absolute value. Frequency
shifts due to electric quadrupole moment, induced polarization and
excess micromotion of the ion depend on electric fields, which are
estimated in this article. Geometry of the trap electrodes also
result in unwanted electric fields which have been considered in
our calculation. Magnetic field induced shift and Stark shifts due
to electro-magnetic radiation at a surrounding temperature are
also estimated. At CSIR-NPL, we are developing a frequency
standard based on the octupole transition for which the systematic
uncertainties are an order of magnitude smaller than that using
the quadrupole transition, as described here.
\end{abstract}

\maketitle
\section{Introduction}
\label{Sec:Introduction}
Recent advances in trapping and laser control of single ion has
started a new era for frequency standards \cite{Wineland_RMP_2013}
in optical frequency region, which can achieve 2-3 orders higher
accuracy and lower systematic uncertainty \cite{Chou_PRL_2010}
than current microwave clocks based on Cesium fountains
\cite{Guena_IEEE_2012, PArora_IEEE_2013}. Realization of more
accurate frequency standards will open up possibilities of vastly
higher speed communication systems and more accurate satellite
navigation systems besides enabling more precise verification of
fundamental physical theories, in particular related to general
relativity \cite{Chou_Science_2010}, cosmology
\cite{SGTuryshev_Cosmology}, and unification of the fundamental
interactions \cite{SGKarshenboim_EPJ_2008}. So far a number of
different ion species have been studied as promising optical
frequency standards at several research institutes worldwide.
These are $^{199}$Hg$^+$ at NIST, USA
\cite{Rosenband_Science_2008}; $^{171}$Yb$^+$ at NPL, UK
\cite{Gill_IEEE_2003} $\&$ PTB, Germany \cite{Tamm_PRA_2009,
Huntemann_PRL_2012}; $^{115}$In$^+$ at MPQ, Germany
\cite{Wang_OC_2007}; $^{88}$Sr$^+$ at NRC, Canada
\cite{Madej_PRA_2013} $\&$ NPL, UK \cite{Margolis_Science_2004};
$^{40}$Ca$^+$ at CAS, China \cite{Gao_CSB_2013} $\&$ NICT, Japan
\cite{Kajita_PRA_2005} and $^{27}$Al$^+$ at NIST, USA
\cite{Chou_PRL_2010}. The accuracy of a frequency standard is
decided by that of the measured atomic transition frequency, which
may shift due to inter-species collisions and their interactions
with external fields. Therefore, it is important to determine
these systematic shifts precisely in order to improve the accuracy
of the realized frequency standard.
At CSIR-NPL, India we are presently developing an optical
frequency standards using Ytterbium-ion \cite{SDe_currentsc_2014}.
It has a narrow $|^2\rm{S_{1/2}; F =0, m_F =
0}\rangle$-$|^2\rm{D_{3/2}; F =2, m_F = 0}\rangle$ quadrupole
transition (E2) and an ultra-narrow $|^2\rm{S_{1/2}; F =0, m_F =
0}\rangle$-$|^2\rm{F_{7/2}; F =3, m_F = 0}\rangle$ octupole
transition (E3) at wavelengths 435.5 nm and 467 nm, respectively
\cite{OurNotation}. These E2 and E3-transitions are at frequencies
$\nu_o = 688\, 358\, 979 \, 309\, 306.62$ Hz \cite{Tamm_PRA_2009}
and $642\, 121\, 496\, 772\, 645.15$ Hz \cite{Huntemann_PRL_2012}
with natural line-widths 3.02 Hz and 1 nHz, respectively. We shall
be probing the E3-transition in our frequency standards. The
nuclear spin $I$=1/2 of $^{171}$Yb$^+$ allows to eliminate the
first-order Zeeman shift. The states associated with the
E3-transition have the highest sensitivity to measure temporal
constancy of fine structure constant and electron-to-proton mass
ratio \cite{Margolis_ConPhys_2010}. In this article we have
estimated five major sources of systematic uncertainties which are
due to the electric quadrupole shift, Doppler shift, dc Stark
shift, black-body radiation shift and Zeeman shift.
%

\section{Trapping of The Ytterbium-ion}
\label{Sec:Trapping of ion}
\begin{figure}
\begin{center}
{
\includegraphics[width=3 in]{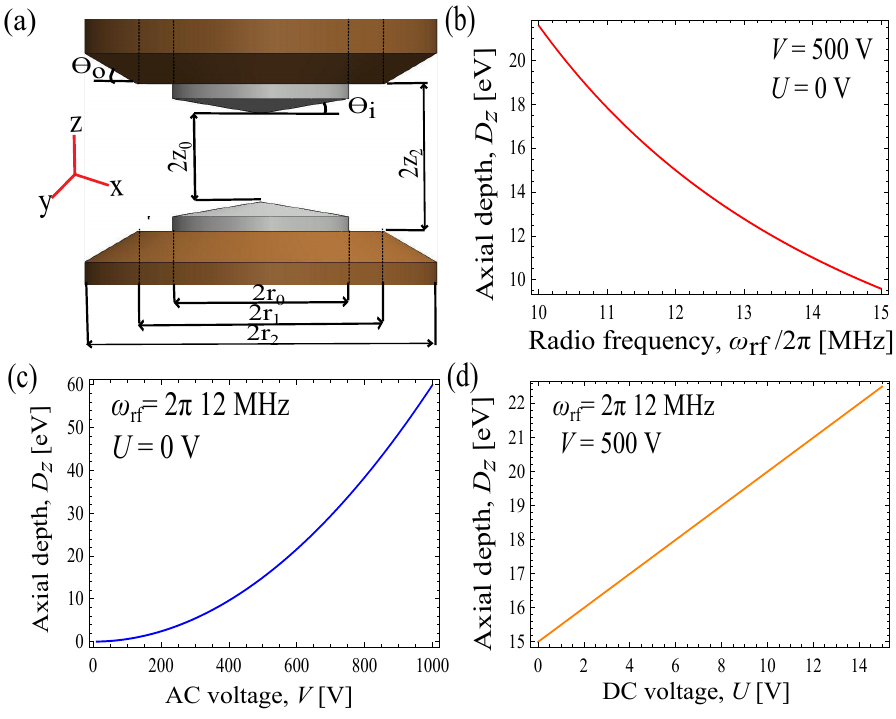}
}\end{center} \caption{(a) The electrode assembly of our end-cap
type Paul trap, where $2z_0 \approx 0.6$ mm, $2z_2 \approx 1.0$
mm, $2r_1 = 1$ mm, $2r_2 = 1.4$ mm, $2r_3 = 2$ mm, $\Theta_i =
10^\circ$ and $\Theta_{o}=45^\circ$. (b), (c) and (d) show axial
trap depth with respect to radio-frequency $\omega_{rf}$, ac $V$
and dc $U$ voltages respectively.}\label{Fig:first}
\end{figure}

A Paul trap \cite{Paul_Rev_1990} of end cap geometry
\cite{Schrama_Optcomm_1993} as shown in Fig.\ref{Fig:first}(a)
will be employed for trapping single $^{171}$Yb$^+$ ion
\cite{SDe_currentsc_2014}. For a pure harmonic trapping potential
$\Phi^{(k=2)}(x,y,z)$ the time dependent trajectory of the ions
\cite{Werth_springer_2010} can be approximated as
\begin{eqnarray}\label{sol-mathieu}
u(t)\approx C\cos\bigg(\beta_{u}
\frac{\omega_{rf}}{2}t\bigg)[1-\frac{q_{u}}{2}\cos(\omega_{rf} t)]
\end{eqnarray}
where $u \in\{x,y,z\}$, $C$ is the amplitude of the motion,
$\omega_{rf}$ is the applied rf, $\beta_{u}=\sqrt{a_{u}+
q_{u}^{2}/2}$, for $a_u$ and $q_u \ll 1$. The stability
parameters, $a_u$ and $q_u$ depend on the applied dc and ac
voltages, respectively. For precision measurements, in a real trap
the anharmonic potential $\Phi^{(k>2)}(x,y,z)$
\cite{SDe_currentsc_2014} are non-negligible. Only the even order
multipoles contribute in the case of a cylindrically symmetric end
cap trap and the dominating perturbation arise from the octupole
term $\Phi^{(k=4)}(x,y,z)$. Neglecting the asymmetries, which may
arise from misalignment of the electrodes and machining
inaccuracies, the trapping potential can be written as,
\begin{eqnarray}\label{quadrupole}
\Phi(x,y,z)&=&\frac{V_{T}(t)}{2{R}^{2}}\bigg[c_{2}(2z^{2}-x^{2}-y^{2})-\frac{c_{4}}{{R}^{2}}(3x^4+3y^4\nonumber\\
&+&8z^4-24x^2z^2-24y^2z^2+6x^2y^2 )\bigg]
\end{eqnarray}
where, $R = \sqrt{{{r_{0}}^2}/2+{z_{0}}^2}$, $V_{T} (t) = U + V
\cos(\omega_{rf}t)$ in terms of the dc component $U$, ac component
$V$ of the trapping voltage and the dimensionless coefficients
$c_{2}$, $c_{4}$ depend on electrode geometry. We have simulated
geometry dependent trap potential using a commercial software
\cite{CPO_USA_2013} and characterized its nature for several trap
geometries as given by Eq.\eqref{quadrupole}. For the trap
geometry shown in Fig. \ref{Fig:first} the coefficients $c_2$ and
$c_4$ have been estimated to be 0.93 and 0.11, respectively. The
restoring force for trapping ions due to $\Phi^{(k=2)}$ produces
an axial trap depth, $D_z(U,V,\omega_{rf}) = U/2 + m{z_{o}}^2
{\omega}^2_{rf} {q_{z}}^2/16 Q$ \cite{Werth_springer_2010} where
$Q$ and $m$ are charge and mass of the ion, respectively. Figure
\ref{Fig:first} (b-d) shows variation of the axial trap depth,
$D_z$ as a function of the control parameters $U$, $V$ $\&$
$\omega_{rf}$ such that $q_z = -16QV c_2/m {\omega}^2_{rf}$ and
$a_z = 32 Q U c_2/m {\omega}^2_{rf}$ lie in the stability region
\cite{Werth_springer_2010}. Throughout this article we have
considered radial coordinate $r$ in the $xy$-plane instead of $x,
\, y$ coordinates, since the trap is axially symmetric.
%

\section{Electric Quadrupole Shift}
\label{Sec:Electric Quadrupole Shift}
Electric quadrupole shift $\Delta \nu_Q$ of the atomic energy
levels is one of the dominating systematic uncertainties for the
precision frequency measurement. It arises due to the interaction
of the atomic quadrupole moment $\Theta(\gamma,J)$ of a state
having spectroscopic notation $\gamma$ and total angular momentum
quantum number $J$ with the external electric field gradient
$\nabla E$, giving a Hamiltonian as
\begin{eqnarray}
H_{Q}=\nabla E \cdot \Theta=\sum^{2}_{q=-2} (-1)^{q}\nabla
E_{q}\Theta_{-q}.
\end{eqnarray}
The quadrupole moment operator $\Theta$ and electric field
gradient $\nabla E$ are tensors of rank two
\cite{Ramsey_oxford_1956}. A non-zero atomic angular momentum
results in a non-spherical charge distribution and the atom
acquires a quadrupole moment. The ground state $\rm{|^{2}S_{1/2};
0, 0\rangle}$ of $^{171}\rm{Yb}^+$ has $\Theta(\rm{S},1/2) = 0$,
but the excited states $\rm{|^{2}D_{3/2}; 2, 0\rangle}$ and
$\rm{|^{2}F_{7/2}; 3, 0\rangle}$ contributes to $\Delta \nu_Q$.
The expectation value of $H_Q$ in reduced form, as given in Ref.
\cite{Itano_Nist_2000}, is
\begin{eqnarray}\label{quadshiftformula}
\langle \gamma J F m_{F}|H_{Q}|\gamma J F m_{F}\rangle &=&
\Theta(\gamma,J) \, \mathcal{F}_Q(I,J,F,m_{F})\nonumber\\
&~& \sum_{q=-2}^{2}\nabla E_{q}D_{0q},
\end{eqnarray}
where $D_{0 q}$ are rotation matrix elements for projecting
components of $\nabla E$ from the principle axes frame that is
defined by the trap axes to the lab frame which is defined by the
quantization direction \cite{Edmonds_princeton_1974} and
\begin{eqnarray}
\mathcal{F}_Q = (-1)^{I+J+F}(2F+1)\left(%
\begin{array}{ccc}
  F & 2 & F \\
 -m_{F} & 0 & m_{F} \\
\end{array}%
\right)\nonumber\\
{\left(%
\begin{array}{ccc}
  J & 2 & J \\
 -J & 0 & J \\
\end{array}%
\right)}^{-1} \left\{
\begin{matrix}{}
  J & 2 & J \\
  F & I & F \\
\end{matrix}\right\}.
\end{eqnarray}
Here the quantities within $( \, )$, $\{ \, \}$ are $3j$,
$6j$-coefficients, respectively and $F$ is total angular momentum
with its projection along the quantization axes $m_{F}$. The
calculated $\mathcal{F}_Q$ for both $\rm{|^{2}D_{3/2}; 2,
0\rangle}$ and $\rm{|^{2}F_{7/2}; 3, 0\rangle}$ states is 1.  Due
to axial symmetry of the trap potential the contributions from
$D_{0 \pm 1}$ cancel with each other and $D_{0 0} = (3
\cos^2\theta -1)/2$, $D_{0 \pm 2} = \sqrt{3/8} \sin^2\theta (\cos
2\phi \mp i \sin 2\phi)$ contribute to Eq. \ref{quadshiftformula},
where $\theta$ and $\phi$ are Euler's angles that rotates the
principle axes frame and overlaps with the lab frame. The tensor
components of $\nabla E$ can be calculated from $E_{x,y,z}$
produced by $\Phi(x,y,z)$ as described in Ref.
\cite{Ramsey_oxford_1956}, which gives $\sum_{q} \nabla
E_{q}D_{0q} = 2 V_T c_2 [D_{00} -D_{02}/\sqrt{6} ]$ and $12 V_T
c_4 [4z^2(D_{00} - D_{02}/\sqrt{6}) - x^2(2D_{00} -
\sqrt{3/2}D_{02}) ] $ for harmonic and anharmonic potentials,
respectively. The measured values of $\Theta(\gamma, J)$ for the
$\rm{|^{2}D_{3/2}; 2, 0 \rangle}$ and $\rm{|^{2}F_{7/2}; 3, 0
\rangle}$ states of $\rm{^{171}Yb^{+}}$ are $2.08(11) e a_{o}^{2}$
\cite{Barwood_PRL_2004} and $-0.041(5) ea_{o}^{2}$
\cite{Huntemann_PRL_2012} respectively, where $e$ is electronic
charge and $a_{o}$ is Bohr radius.
\begin{figure}
\begin{center}{
\includegraphics[width=3 in]{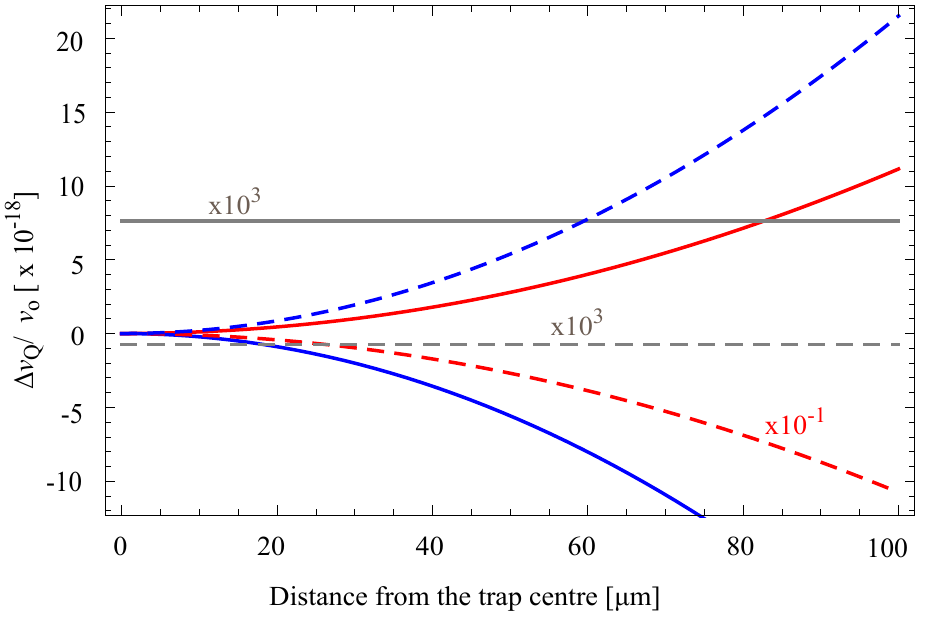}
}\end{center} \caption{(Color online) Spatial dependence of the
fractional electric quadrupole shifts $\Delta \nu_Q/\nu_o$ at the
E2 and E3-clock transitions of $\rm{^{171}Yb^+}$, which are
distinguished by solid and dashed lines, respectively. The
quadrupole trapping potential produces a constant shift (gray) and
the spatial dependence along the radial (red) and axial (blue)
directions arise from the anharmonic components \cite{Figure}.
}\label{Fig:Qshift}
\end{figure}
The harmonic component of the trapping potential gives a constant
electric field gradient however a spatial dependence comes from
the anharmonic component, which introduces an uncertainty in the
measured $\Delta \nu_{Q}$ due to motion of the ion. We estimate
the quadrupole shift due to $U$ since the contribution from the
$\rm{rf}$ averages to zero for first order electric quadrupole
shift and for second order it is zero in case of $\rm{^{171}Yb^+}$
\cite{Schneider_PRL_2005}. Figure \ref{Fig:Qshift} shows the
estimated fractional quadrupole shifts $\Delta \nu_{Q}/\nu_{0}$
due to $\Phi^{(2)}$ and $\Phi^{(4)}$ for the E2 and E3 -
transitions of $\rm{^{171}Yb^{+}}$ respectively as a function of
the radial and axial distance from the trap center. The shifts due
to $\Phi^{(4)}$ computed for $\emph{U}=10$ V and $\theta =
0^\circ$ are found to be three orders of magnitude smaller than
the contribution due to $\Phi^{(2)}$, which are $\approx~5.25$ Hz
and $\approx-0.51$ Hz for E2 and E3-transitions respectively and
have no spatial dependence. The frequency shift can be cancelled
in different ways as described in Ref. \cite{Madej_PRA_2013},
which could be opted depending on the system. The magnitude of the
quadrupole shift is twice along the $z$-axis than they are along
the $x, \, y$-axes but in opposite directions, respectively. We
shall measure $\Delta \nu_Q$ separately by quantizing the ion
along three mutually orthogonal directions of the principle axes,
\emph{i}. \emph{e}. $\theta = 0^\circ$, using magnetic fields of
equal amplitude. Averaging these three would eliminate the total
quadrupole shift \cite{Itano_Nist_2000}.
%

\section{Doppler Shift}
\label{Sec:Doppler Shift}
The relative motion between the laboratory and the ionic frames of
reference introduces a shift in the observed frequency. The
absorbed or emitted radiation $E_o \cos (\omega_o t)$ at frequency
$\omega_o = 2\pi \nu_o$ (wavelength $\lambda_o$) experiences a
phase modulation $\eta \sin \omega_s t$ due to secular motion of
the trapped ion at frequency $\omega_s$. The modulation depth
$\eta = \Delta\omega_o/\omega_s$ depends on the Doppler shift
$\Delta \omega_{o} = 2\pi v/ \lambda_{o}$ due to ion's velocity $v
= \omega_s r$. A modulated spectrum $E_o \cos (\omega_o t) \pm
\eta E_o \cos(\omega_o \pm \omega_s) t$ is expected when the ion
is confined within $r < \lambda_o$ \cite{Dicke_PR_1953}, which is
generally observed in an absorption spectroscopy for a narrow
transition \cite{Berguist_PRA_1987}. This allows accurate
determination of the first order Doppler unshifted $\nu_o$ for a
laser cooled ion. However the second order Doppler effect
introduces a frequency shift, which is given by
\begin{equation}\label{seconddoppshift}
\frac{\Delta\nu_{D2}}{\nu_o} = -\frac{v^2}{2c^2} = \frac{2
\varepsilon_k}{mc^2}
\end{equation}
for kinetic energy $\varepsilon_k$ of the ion; $c$ is speed of
light. For a laser cooled ion at 1 mK the fractional frequency
uncertainty due to the temperature dependent second order Doppler
effect is $\approx 10^{-19}$.
\begin{figure}
\begin{center}
{\includegraphics[width=3 in]{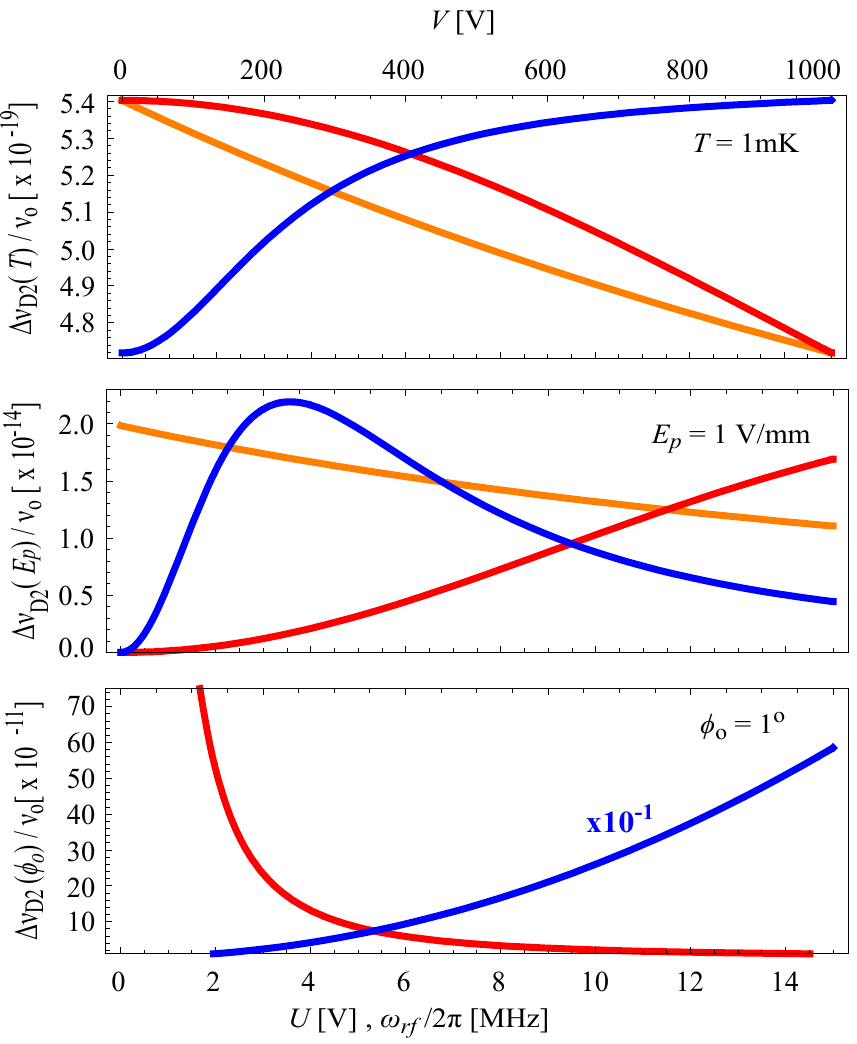}}
\end{center}
\caption{ (Color online)Variation of the second order Doppler
shifts with respect to the trap parameters - radio-frequency
$\omega_{rf}$ (red), ac $V$ (blue) and dc $U$ (orange) voltages
resulting from (a) temperature of the ion $T = 1$ mK, (b) patch
potential that ion experiences $E_p = 1$ V/mm and (c) relative
phase difference $\phi_o = 1^\circ$ of the rf at two electrodes
\cite{Figure}.}\label{Fig:dopp2}
\end{figure}
Velocity of the trapped ion can be calculated from its trajectory,
which gets deviated from Eq.\eqref{sol-mathieu} due to slowly
varying stray electric fields. This can result from the patches of
unwanted atoms on the electrode surface and relative phase
differences of the rf on them. Over the time, Tantalum electrodes
get coated with $^{171}\rm{Yb}$ atoms coming out of the oven. The
differential work-function of Ytterbium and the Tantalum results
in an electric field $\vec{E}_p$, which varies slowly with the
deposition of atoms. As a result of an extra force, $Q \vec{E}_p$,
the minimum of the confining potential shifts by an amount $C_o =
Q \vec{E}_p \cdot\hat{u}/(m \omega_{s}^2)$ and the micromotion
increases \cite{Berkeland_JAP_1998}. A difference in path lengths
and non-identical dimensions of the electrodes introduce a phase
difference $\phi_o$ between the rf on the electrodes as $V
\cos(\omega_{rf}t \pm \phi_o/2)$. For small $\phi_o$, \emph{i}.
\emph{e}., $\sin \phi_o = \phi_o$ one can approximate this as two
parallel plates separated by $2 z_o/\alpha$ and at potentials $\pm
V \phi_o \sin (\omega_rf t)/2$ which are subjected in addition to
the rf, where the geometric factor $\alpha \approx 0.8$ for trap
geometries satisfying $r_o^2 = 2 z_o^2$
\cite{Gabrielse_PRA_1984,Beaty_PRA_1986}. For our trap geometry
this generates an extra ac electric field $V \phi_o \alpha/2 z_o
\times \sin (\omega_{rf}t) \hat{z}$ which increases micromotion
along the axial direction.
\begin{figure}
\begin{center}
 {\includegraphics[width=3
in]{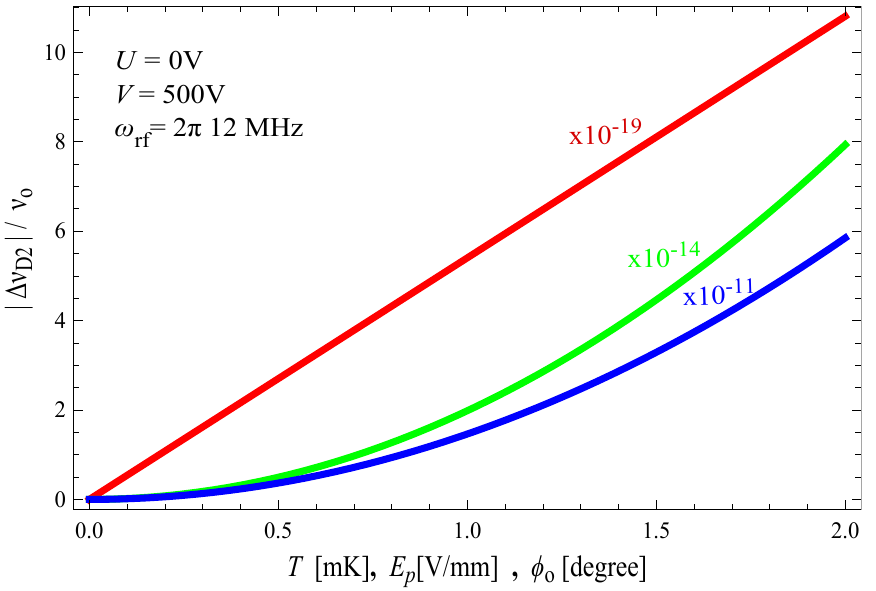}}
\end{center}
\caption{ (Color online) Fractional second order Doppler shift
$\Delta \nu_{2D}/\nu_o$ with: temperature $T$ (red); patch
potential generated electric field $E_p$ (green) and phase
difference $\phi_{o}$ of rf at the electrodes (blue)
\cite{Figure}.}\label{Fig:dopp1}
\end{figure}
Additional electric fields subject ion to excess force hence the
ion trajectory gets modified as
\begin{eqnarray}
u(t)&\cong&[C_{0}+C\cos(\omega_{rf}t)]\bigg[1+\frac{q_{u}}{2}\cos(\omega_{rf}
 t)\bigg]\nonumber\\
 &-&\frac{1}{4}q_{u}z_{0}\alpha \phi_{0}\sin(\omega_{rf}
 t)\delta_{u,z}.
\end{eqnarray}
where $\delta_{u,z}$ is Kronecker delta. This gives excess kinetic
energy to the ion as described in Ref. \cite{Berkeland_JAP_1998}.
The average kinetic energy of the ion is
\begin{eqnarray}\label{kinetic3}
\varepsilon_{k,u}&=&\frac{1}{4}m
C^{2}\bigg[\omega_{s}^{2}+\frac{1}{8}q_{u}^{2}{\omega_{rf}}^{2}\bigg]+\frac{4}{m}\bigg[\frac{Qq_{u}
\vec{E}_p \cdot
\hat{u}}{(2a_{u}+q_{u}^{2})\omega_{rf}}\bigg]^{2}\nonumber\\
&+&\bigg[\frac{m(q_{u}z_{0}\alpha\phi_{o}\omega_{rf})^{2}}{64}\bigg]\delta_{u,z}
\end{eqnarray}
where the first term depends on temperature of the ion and
remaining two terms are due to the $\vec{E}_p$ and $\phi_o$,
respectively. The fractional frequency shift which is independent
of $\nu_o$ can be calculated using Eq.\eqref{seconddoppshift}.
Each component of $\Delta\nu_{D2}/\nu_o$ depends on the trap
parameters $\omega_{rf}$, $V$ and $U$ which are shown in Fig.
\ref{Fig:dopp2}. Figure \ref{Fig:dopp1} shows that
$\Delta\nu_{2D}/\nu_o$ due to the patch potentials and ac phase
difference at the two counteracting electrodes can produce orders
of magnitude larger frequency shift than any other systematic
effects. These are also discussed by Berkeland \emph{et}.
\emph{al}. in Ref. \cite{Berkeland_JAP_1998} and by P. Gill in
Ref. \cite{Gill_Metrologia_2005}. This concludes in order to build
a frequency standard with fractional accuracy $~ 10^{-17}$, one
has to control $\phi_o$ at a level $\approx 0.5$ milli-degree and
$E_p < 20$ mV/mm respectively. We shall employ two additional
pairs of counteracting electrodes in the radial plane for
cancelling stray potentials that ion experiences and the accurate
machining will be essential for maintaining nearly zero path
difference of the applied rf to the electrodes.
%

\section{DC Stark shift}
\label{Sec:Stark shift}
Interaction of electric dipole moment (EDM) of an atom with an
electric field results in Stark shift \cite{stark,Itano_Nist_2000}
of the atomic energy levels. The interaction energy is given as
\begin{equation} \label{eq-schro}
H_I=-\vec{E} \cdot \vec{d},
\end{equation}
where $\vec{E}$ is the electric field and $\vec{d}$ is the
electric-dipole operator. As describe in Sec. \ref{Sec:Doppler Shift} in a
real experiment the patch potentials lead to non zero dc electric fields at
ion's location and can introduce dc Stark shift. The
electro-magnetic (EM) radiations at the non-zero temperature of
the apparatus also introduces dc Stark shift which is known as
black body radiation (BBR) shift. For $^{171}$Yb$^+$ the first
order Stark shift is zero because ion acquires a zero permanent
EDM. The coupling of the $^2\rm{S}_{1/2}$, $^2\rm{D}_{3/2}$ and
$^2\rm{F}_{7/2}$ states in $^{171}$Yb$^+$ to all the other states
via electric dipole interaction results to a non-zero second-order
Stark shift which is not negligible. An induced EDM produces second
order Stark shift \cite{stark} as
\begin{equation}\label{eq-8}
\Delta \nu_{dc} = -\frac{1}{2h}\alpha E_p^2,
\end{equation}
where $h$ is the Plank constant, polarizability $\alpha$ has both
scalar ($\alpha_{0}$) and tensor ($\alpha_2$) contributions. The
effective $\Delta \nu_{dc}$ is calculated as the difference
between the shifts of the states involved in the clock transition
\cite{Itano_Nist_2000,fieqst} as
\begin{equation}\label{eq-stark}
\Delta \nu_{dc}  = \frac{1}{4h}\left[ 2 \Delta \alpha_0 + \Delta
\alpha_2(3\cos^2\theta-1)\right] E_p^2,
\end{equation}
where $\Delta \alpha_0$ and $\Delta \alpha_2$ are the
polarizability differences of the states associated with the clock
transition. The Stark shift becomes independent of $\Delta
\alpha_2$ at $\theta=54.73^\circ$ but in our experiment $\theta
\approx 0^\circ$ fixed by geometry of the apparatus. Here we estimate
the second order Stark shifts resulting from dc electric field and
EM-radiation.
\begin{figure}
\begin{center}
 {\includegraphics[width=3
in]{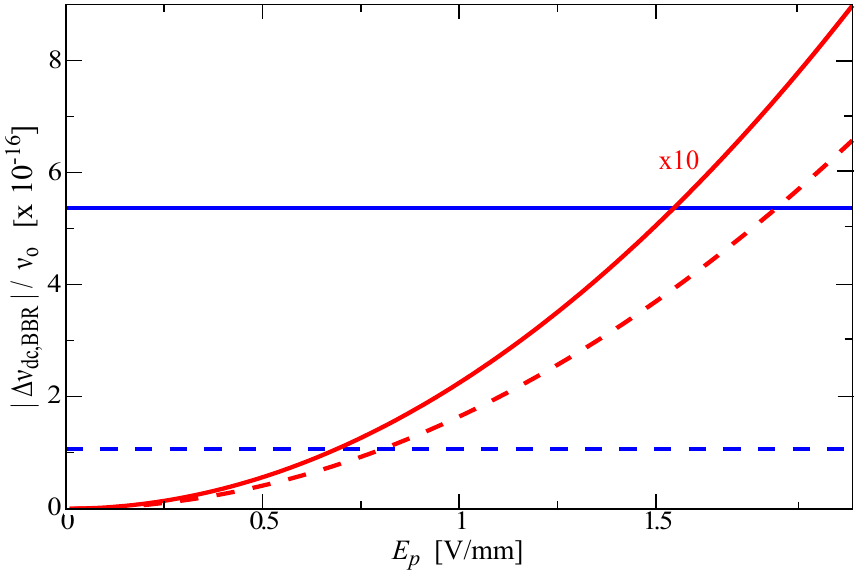}}
\end{center}
\caption{ (Color online) Variation of the fractional dc Stark shift
$\Delta\nu_{dc}/\nu_o$ with electric field $E_p$ (red) and the fractional BBR shift $\Delta
\nu_{BBR}/\nu_o$ at 300 K (blue). E2 and E3-transitions are distinguished by
solid and dashed lines, respectively
\cite{Figure}.}\label{Fig:Stark}
\end{figure}
The Stark shift due to the $\Delta\alpha_2$ vanishes at the ground
state because of its symmetric nature but not for the
$^2$D$_{3/2}$ and $^2$F$_{7/2}$ states. Using the measured
polarizibilities of the $^2$S$_{1/2}$, $^2$D$_{3/2}$
\cite{Schneider_PRL_2005} and $^2$F$_{7/2}$
\cite{Huntemann_PRL_2012} states variation of $\Delta
\nu_{dc}/\nu_o$ with $E_p$ for the E2 and E3-transitions are shown
in Fig. \ref{Fig:Stark}.
The electric field associated with the EM-radiation produced due to finite
temperature of the apparatus and particularly the oven producing
an atomic beam gives rise to BBR shift. The
temperature dependent electric and magnetic fields are given by the Planck's law \cite{bbr3}
as
\begin{equation}\label{eq-3}
E^2(\omega)d\omega = B^2(\omega)d\omega = \frac{8\alpha^3}{\pi}
\frac{\omega^3d\omega}{\exp(\frac{\omega}{k_BT})-1},
\end{equation}
where $B$ is magnetic field, $\alpha$ is emissivity of the
material and $\omega$ is frequency of EM-radiation. The wavelength
corresponding to the maximum of the spectral energy density at 300
K is 9.7 $\mu$m \cite{microm}, which is large compared to the
longest transition wavelength $\approx$ 2.4 $\mu$m in
$^{171}$Yb$^+$. Therefore to a good approximation the BBR generated RMS
amplitude of $E$ and $B$ fields are written as $\langle E^2 \rangle = E_o^2 \times
(T/300)^4$ and $\langle B^2 \rangle = B_o^2 \times (T/300)^4$, where $E_o
= 831.9$ V/m and $B_o = 2.775 \times10^{-6} $ T, respectively
\cite{bbreq}. The magnetic field contributes to a Zeeman shift,
which will be discussed in the section \ref{Sec:Zeeman Shift}. The
contribution due to $\alpha_2$ can be neglected for an isotropic
EM-radiation and the effective BBR shift can be written as
\begin{equation}
\Delta \nu_{BBR} = -\frac{1}{2h}\Delta\alpha_0 \,
E_o^2\left(\frac{T}{300}\right)^4.
\end{equation}
At room temperature a shift of about 0.36 Hz and 0.068 Hz are
estimated for the E2 and E3-transitions, respectively.
%

\section{Zeeman Shift}
\label{Sec:Zeeman Shift}
Zeeman shift arises due to the interaction of atomic and nuclear
magnetic moments $\mu_J$ and $\mu_I$ with an external magnetic field. In an
experiment, magnetic field appears from the BBR, geomagnetic and
stray fields. The E2 and E3-clock transitions are insensitive to
the linear Zeeman effect since ground and excited states
have $m_F=0$ states associated with them. Since the nuclear $g$-factor $g_I$ is
much smaller than the electronic $g$-factor $g_J$, the second order
Zeeman shift \cite{zeeman} of the sublevels can be approximated only in terms of $g_J$ as
\begin{eqnarray}\label{eq-13}
\Delta \nu_{\rm{QZ}} = -\left(\frac{g_JeB}{4\pi m}\right)^2
\sum_{F'}\frac{|\mathcal{F}_Z(I,J,F,F',m_{F})|^2}{\Delta
\nu_{HFS}}, \nonumber\\
\end{eqnarray}
where $\Delta \nu_{HFS}$ is the hyperfine splitting of the states
and the matrix element $\mathcal{F}_Z = \langle F', m'_F|J_{z}|F, m_F \rangle$
\cite{zeeman1} is given as
\begin{eqnarray}\label{eq-14}
               \mathcal{F}_Z &=& \sqrt{I(I+1)(2I+1)(2F+1)(2F'+1)} \nonumber \\
                                    &~& \left( \begin{array}{ccc}
                                            F & 1 & F'\\
                                            -m_F & 0 & m_F
                                           \end{array}\right) \left\{ \begin{array}{ccc}
                                            I & F & J\\
                                            F' & I & 1
                                           \end{array}\right\}.
\end{eqnarray}
The calculated $|\mathcal{F}_Z|^2 = 1/4$ for the $^2$S$_{1/2}$,
$^2$D$_{3/2}$ and $^2$F$_{7/2}$ states in $^{171}$Yb$^+$. Their
$g_J$ values are 1.998, 0.8021, 1.1429 and $\Delta \nu_{HFS}$ are
12.643 GHz, 0.86  GHz, 3.62 GHz, respectively
\cite{NIST_database}. The geomagnetic field in New Delhi, India is
approximately $50~\mu$T which produces $\Delta \nu_{\rm{QZ}}$ of
38.75 Hz, 91.15 Hz, and 44.19 Hz at the $^2$S$_{1/2}$,
$^2$D$_{3/2}$ and $^2$F$_{7/2}$ states, respectively. This results
in a net second order Zeeman shift of 52.40 Hz and 5.44 Hz for the
E2 and E3-clock transition, respectively. These are much larger
than the shift produced by the magnetic field of the BBR at the
room temperature whose values are $0.16$ Hz and $0.017$ Hz for the
E2 and E3-transitions, respectively.
%

\section{Conclusion}
\label{Sec:Conclusion}
\begin{table}[h]
\caption{Fractional shifts due to the systematic effects for the
E2 and E3-transitions. The shifts are estimated at the room
temperature $T=300$ K, rf phase difference $\phi_o=0.5$
mili-degree, stray electric and magnetic fields $E_p=20$ mV/mm and
$B=1$ $\mu$T. Numerical values of $\phi_o$, $E_p$ and $B$ which
are used here are typical values that can be achieved with proper
minimization techniques.} \label{Tab:Uncertainty}
\begin{tabular}{lll}
\hline
Systematic                & E2-transition          & E3-transition          \\
effect                    & [$\times 10^{-17}$]    & [$\times 10^{-18}$ ]   \\
\hline
Electric quadrupole       & $762$                  & $-789$                 \\
Second order Doppler      & $-1$                   & $-1$                   \\
dc Stark                  & $-0.09$                & $-0.07$                \\
BBR: dc Stark             & $-52.3$                & $-106$                 \\
Second order Zeeman       & $-3.04$                & $-3.39$                \\
BBR: second order Zeeman  & $-23.2$                & $-26.4$                \\

\hline
\end{tabular}
\end{table}
The systematic shifts from different source have been estimated
for the E2 and E3-transitions of $^{171}$Yb$^+$ and summarized in
Tab. \ref{Tab:Uncertainty}. Even though the electric quadrupole
shift is the largest, averaging the measured frequency along three
orthogonal directions effectively cancels $\Delta \nu_Q$. Three
pairs of Helmholtz coils will be installed for defining the
quantization axes. These coils will be used to cancel the static
stray magnetic fields as well, for minimizing the quadratic Zeeman
shift. The thermal part of $\Delta\nu_{D2}(T)$ is an order of
magnitude smaller compared to the frequency standard that we aim
for. Careful wiring for supplying rf and accurate machining of the
electrodes is very important for making $\phi_o \approx 0^\circ$.
Two pairs of electrodes will be installed in the radial plane for
compensating the local electric fields that a trapped ion feels,
which is required for minimizing $\Delta\nu_{D2}(E_p)$ and
$\Delta\nu_{dc}$. Surrounding temperature at the position of ion
needs to be measured accurately \cite{Bloom_Nature_2014} for
estimating the Stark and Zeeman shifts produced by BBR, which is
the dominating systematic effect (Tab. \ref{Tab:Uncertainty}).
From our estimation, the E3-transition can provide an order of
magnitude accurate frequency standards than the E2-transition of
$^{171}$Yb$^+$.


\begin{thebibliography}{100}
\bibitem{OurNotation}{Throughout this manuscript, the atomic energy state total angular momentum $F$ and
projection of it along the quantization axes defined by the
applied magnetic field $m_{F}$ are contained within the ket.}
\bibitem{Figure}{Multiplication factors, as indicated in the fugures, are used for accomodating multiple plots in a
single figure.}
\end{thebibliography}
\end{document}